\documentclass{article}
\title{Search strategies for developing characterizations of graphs without small wheel subdivisions}
\author{{Rebecca Robinson and Graham Farr}\\
	[1ex] Clayton School of Information Technology\\
	Monash University\\
	Clayton, Victoria, 3800\\
	Australia\\[1ex]
	Rebecca.Robinson@monash.edu\\
	Graham.Farr@monash.edu
	}
\date{April 30, 2009}

\usepackage[dvips]{graphicx}
\usepackage{amsmath, amsthm, amssymb}
\usepackage[toc,page]{appendix}

\begin{document}
\maketitle

\begin{abstract}
Practical algorithms for solving the Subgraph Homeomorphism Problem are known for only a few small pattern graphs: among these are the wheel graphs with four, five, six, and seven spokes. The length and difficulty of the proofs leading to these algorithms increase greatly as the size of the pattern graph increases. Proving a result for the wheel with six spokes requires extensive case analysis on many small graphs, and even more such analysis is needed for the wheel with seven spokes. This paper describes algorithms and programs used to automate the generation and testing of the graphs that arise as cases in these proofs. The main algorithm given may be useful in a more general context, for developing other characterizations of SHP-related properties.
\end{abstract}

\section{Introduction}
\label{intro}

The \emph{Subgraph Homeomorphism Problem}, SHP --- also known as \emph{topological containment} --- is an important problem in graph theory, and belongs to Garey and Johnson's original list of NP-complete problems \cite{G&J79}. Any fixed \emph{pattern graph} $H$ gives rise to the problem:

\begin{tabular}{l}
\\
SUBGRAPH HOMEOMORPHISM ($H$) (abbreviated SHP($H$))\\
Instance: Graph $G$.\\
Question: Does $G$ contain a subdivision of $H$?\\
\\
\end{tabular}

It is known that this problem can be solved in polynomial time for any fixed pattern graph \cite{R&S95}, but practical algorithms exist only for a few small pattern graphs. Among these are certain members of the wheel class of graphs, for which characterizations have been obtained: $W_{4}$ and $W_{5}$ in \cite{Farr88}, and $W_{6}$ in \cite{Robinson08}. A result has also been obtained for $W_{7}$ in \cite{Robinson09}, which leads to an efficient algorithm for solving SHP($W_{7}$). The length and difficulty of the proof increases for each $W_{k}$ as $k$ increases. The $W_{4}$ proof takes only a paragraph, and the $W_{5}$ proof occupies 7 pages. The $W_{6}$ proof, however (16 pages, with some automated analysis), requires extensive amounts of repetitive case analysis, and the $W_{7}$ proof even more so (around 90 pages, also with automated analysis). This case analysis involves looking at numerous small graphs of bounded size, and searching for $W_{k}$-subdivisions in those graphs.

This paper presents some algorithms developed to automate parts of the searching and analysis required in developing the results for $W_{6}$- and $W_{7}$-subdivisions. The proofs of these results are similar in structure: both involve beginning with a pattern graph for which some good characterization already exists, then examining all possible ways in which certain structures can be added to this graph to satisfy some necessary condition. It must then be determined whether or not the resulting graphs topologically contain the pattern graph for which the new characterization is desired. This technique involves testing many small graphs for the presence of $W_{6}$- or $W_{7}$-subdivisions.

Since the process of constructing these small graphs is repetitive in nature, it was possible to create a program that automates their construction. Given the sheer number of test cases that arise, particularly for the $W_{7}$ result, this program is important in obtaining the information necessary to complete these proofs, as examining each graph individually by hand would take an inordinate amount of time. In particular, one of the key algorithms used in the program (given in Section \ref{furthertests}) could be applicable in a broader context --- most obviously for developing characterizations relating to wheels with more than seven spokes, but also potentially for obtaining results for subdivisions of graphs other than wheels, if similar techniques can be used.

Each of the graphs generated by the program must be individually tested for the presence of a $W_{6}$- or $W_{7}$-subdivision. Those that do not contain such a subdivision require further analysis in the proof, and so are given as output. In order to successfully perform a test for the presence of a $W_{k}$-subdivision on each generated graph, an algorithm is required that will solve SHP($W_{k}$) for each graph. We used a naive algorithm which runs in exponential, rather than polynomial time. It performs adequately for the small input graphs that arise in the proofs, and its correctness is easily verifiable. 
 
Section \ref{testcases} describes the types of case analysis required in the proofs of \cite{Robinson08} and \cite{Robinson09}, gives the algorithms that have been developed for generating these cases, and demonstrates how these algorithms are used in the context of the proofs. Section \ref{wk_alg} gives the exponential-time algorithm for solving SHP($W_{k}$) that is used in testing the generated graphs. Each algorithm mentioned has been implemented in C, and the code can be found in the Appendices at the end of the paper. The complete code for all implementations can be found online at \verb|http://www.csse.monash.edu.au/~rebeccar/wheelcode.html|.

\section{Definitions}
\label{defs}

If $W$ is a subset of graph $G$, then $G|W$ denotes the set of all maximal subsets $U$ of $V(G)$ such that any two vertices of $U$ are joined by a path in $G$ with no internal vertex in $W$. Each element of $G|W$ is referred to as a \emph{bridge} of $G|W$.

A vertex $v$ of degree 2 is \emph{contracted} in a graph $G$ by adding an edge between $v$'s neighbours, if such an edge does not already exist, then deleting $v$.

\section{Automated generation of test cases}
\label{testcases}

In developing proofs for the characterizations of \cite{Robinson08} and \cite{Robinson09}, algorithms were written to generate specific graphs that arise as cases in these proofs, then test these graphs for the presence of a $W_{6}$- or $W_{7}$-subdivision. This section outlines how such automated graph generation is done. 

Section \ref{wheelproof} describes the \texttt{wheelproof} function, which is used to perform preparatory work in the proofs of the $W_{6}$ and $W_{7}$ results. Its role in the proofs is simple, but it provides a good illustration of the search techniques used. Section \ref{furthertests} describes the \texttt{exception\_generator} function, which is a more general function that is applicable in a wider range of situations, and as such it is used often throughout the proofs.

\subsection{\texttt{wheelproof(k)}: Initial generation of `exception' graphs}
\label{wheelproof}

Each of the proofs for the theorems regarding graphs with no $W_{k}$-subdivision, for $5 \le k \le 7$, follows a similar overall structure:

\begin{itemize}
\item  Firstly, it is proved that for some graph $G$ that meets the conditions of the hypothesis, there must exist some $W_{k-1}$-subdivision $H$ centred on a specific vertex $v_{0}$ of degree $\ge k$.
\item It is then observed that some neighbour $u$ of $v_{0}$ exists such that $u$ is not a neighbour of $v_{0}$ in $H$, and that, since $G$ is 3-connected, there must be two disjoint paths $P_{1}$ and $P_{2}$ in $G$ from $u$ to $H$ that do not meet $v_{0}$.
\item All possible placings of the paths $P_{1}$ and $P_{2}$ must be examined, and each resulting graph must contain a $W_{k}$-subdivision, if it is to satisfy the hypothesis of the Theorem. In situations where the graph $H\cup P_{1}\cup P_{2}$ contains a $W_{k}$-subdivision, this is simple. Where this is not the case, closer examination of the structure of $G$ is required.
\end{itemize}

The function \texttt{wheelproof(k)} was created specifically to generate all possible placings of $P_{1}$ and $P_{2}$ for which $H\cup P_{1}\cup P_{2}$ does not contain a $W_{k}$-subdivision (for any input $k$). We refer to such graphs as \emph{exception} graphs.

This function firstly constructs the graph $W_{k-1}$, then generates all possible ways in which a $k$\textsuperscript{th} neighbour can be added to the centre vertex, while still preserving the 3-connectivity of the graph. For each graph $G_{i}$ that is generated, the function \texttt{findkwheel} is then run with the arguments $(G_{i}, k)$, to test for the presence of a $W_{k}$-subdivision. Any graph generated which is found not to contain such a subdivision is recorded as an exception graph; the function returns a list of all such graphs found.

The C implementation for \texttt{wheelproof(k)} is given in Appendix \ref{app_wheelproof}.

Running the \texttt{wheelproof} function with an input of $k = 4$ generates no exception graphs. This is to be expected, as the characterization for graphs topologically containing $W_{4}$ is as follows \cite{Farr88}:

\begin{quotation}
If $G$ is a 3-connected graph and $v$ is any vertex of degree $\ge 4$ in $G$, then $G$ contains a $W_{4}$-subdivision centred on $v$.
\end{quotation}

\begin{figure}[!hpt]
\begin{center}
\includegraphics[width=0.8\textwidth]{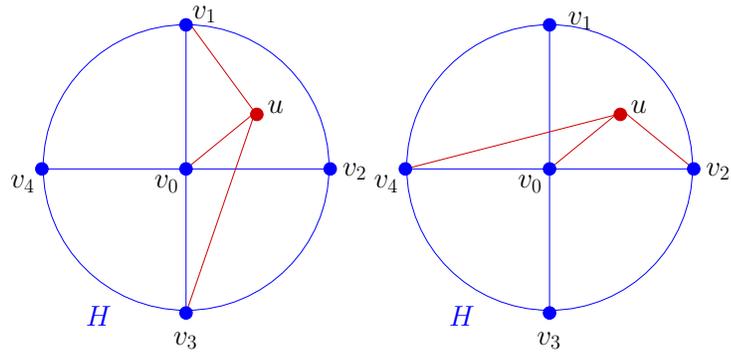}
\caption[wheelproof(5) output]{Exception graphs generated from \texttt{wheelproof(5)} output. They are isomorphic.}
\label{W5_exceptions}
\end{center}
\end{figure}

The output of \texttt{wheelproof(5)} is also as expected, returning two different exception graphs (shown in Figure \ref{W5_exceptions}), each of which is isomorphic to the starting graph of Subcase 2b in Theorem 3 of \cite{Farr88}. (This theorem characterizes graphs containing no $W_{5}$-subdivision; the subcase mentioned deals specifically with a section of the proof requiring the imposition of extra restrictions on the input graph $G$, namely, that $G$ contains no internal 3-edge-cutsets, and that $G$ contains a cycle of length at least 5 disjoint from the selected vertex of degree $\ge 5$.)

\begin{figure}[!hpt]
\begin{center}
\includegraphics[width=0.4\textwidth]{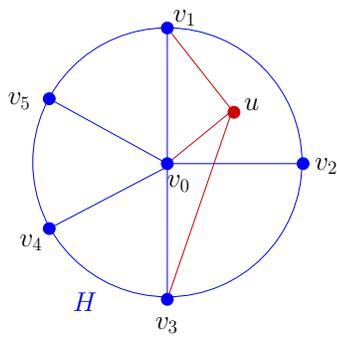}
\caption[wheelproof(6) output]{Exception graph isomorphic to the five graphs generated from \texttt{wheelproof(6)} output.}
\label{W6_exceptions}
\end{center}
\end{figure}

The output of \texttt{wheelproof(6)} generates five different exception graphs. These graphs are isomorphic, and thus further analysis of only one is sufficient (shown in Figure \ref{W6_exceptions}). Such analysis is given in Case (b)(ii) of the main theorem of \cite{Robinson08}.

\begin{figure}[!hpt]
\begin{center}
\includegraphics[width=\textwidth]{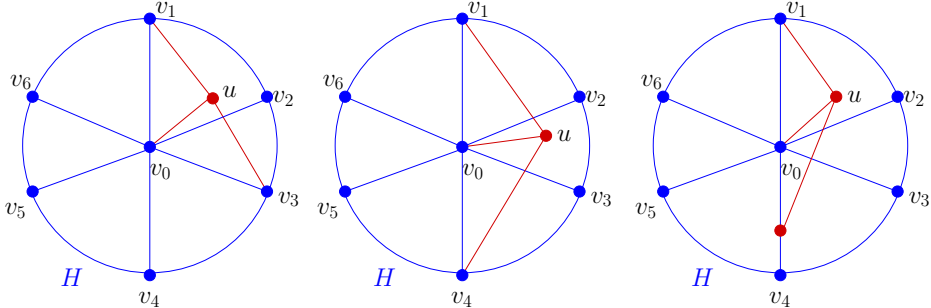}
\caption[wheelproof(7) output]{Exception graphs for $W_{7}$ case. Each of the 15 graphs generated from the output of \texttt{wheelproof(7)} is isomorphic to one of these three graphs.}
\label{W7_exceptions}
\end{center}
\end{figure}

The output of \texttt{wheelproof(7)} gives 15 different graphs, but when examined for isomorphism, this number is reduced to three (see Figure \ref{W7_exceptions}). Each of these three graphs are analysed further in Cases (b)(i), (b)(ii), and (b)(iii) of the main theorem of \cite{Robinson09}.

\subsection{Further automation used in proofs}
\label{furthertests}

Certain other situations arise in the proofs characterizing the $W_{6}$ and $W_{7}$ cases which lend themselves to further automated generation of test cases. These situations all have the following features:

\begin{itemize}
\item Only part of the structure of $G$ is known, represented by a smaller graph, $G'$. Each edge in $G'$ corresponds to a path in $G$.
\item $G'$ contains a separating set $S$, and $G'$ contains a number of bridges of $G'|S$.
\item It is unknown whether $S$ is also a separating set of $G$, or if each bridge of $G'|S$ is contained in a separate bridge of $G|S$.
\end{itemize}

The proof requires that it be known how many bridges of $G'|S$ are contained in separate bridges of $G|S$. Thus, a path $P$ is added to $G'$, where $P$ is disjoint from $G'$ except at its endpoints, each of which are in two separate bridges of $G'|S$ (but not in $S$). All possible graphs $G'\cup P$ are generated, for all possible placements, in $G'\setminus S$, of the endpoints of $P$. Each generated graph $G'\cup P$ is then tested for the presence of a $W_{k}$-subdivision, and only those graphs which do not contain such a subdivision require further analysis.

The function \texttt{exception\_generator} is used to automate this process. This function takes a graph $G$, and the vertex sets of two subgraphs of $G$, say $A$ and $B$. The function generates all possible graphs of the form $G\cup P$, where $P$ is some path disjoint from $G$ except at its endpoints, one of which is in $A$ and one of which is in $B$. The function tests each generated graph for the presence of a $W_{k}$-subdivision, and outputs those that do not contain such a subdivision.

An outline of the algorithm is as follows:

\begin{itemize}
\item[] For each pair of vertices $i, j$, where $i\in V(A)$ and $j\in V(B)$:
	\begin{itemize}
	\item[] Add edge $ij$
	\item[] Check for existence of $W_{7}$-subdivision
	\item[] Remove edge $ij$
	\item[] For each vertex $k$ in $V(A)$ adjacent to $i$:
		\begin{itemize}
		\item[] Create a new vertex $v$, and subdivide the edge $ik$ into two new edges $iv$ and $vk$
		\item[] Add edge $vj$
		\item[] Check for existence of $W_{7}$-subdivision
		\item[] Remove edge $vj$
		\item[] Contract vertex $v$
		\end{itemize}
	\item[] For each vertex $k$ in $V(B)$ adjacent to $j$:
		\begin{itemize}
		\item[] Create a new vertex $v$, and subdivide the edge $jk$ into two new edges $jv$ and $vk$
		\item[] Add edge $vi$
		\item[] Check for existence of $W_{7}$-subdivision
		\item[] Remove edge $vj$
		\item[] For each vertex $l$ in $V(A)$ adjacent to $i$:
			\begin{itemize}
			\item[] Create a new vertex $u$, and subdivide the edge $il$ into two new edges $iu$ and $ul$
			\item[] Add edge $vu$
			\item[] Check for existence of $W_{7}$-subdivision
			\item[] Remove edge $vu$
			\item[] Contract vertex $u$
			\end{itemize}
		\item[] Contract vertex $v$
		\end{itemize}
	\end{itemize}
\end{itemize}

The implementation of this algorithm is given in Appendix \ref{app_exception_generator}.

\subsection{Using \texttt{exception\_generator}: an example}
\label{example}

\begin{figure}[!hpt]
\begin{center}
\includegraphics[width=0.6\textwidth]{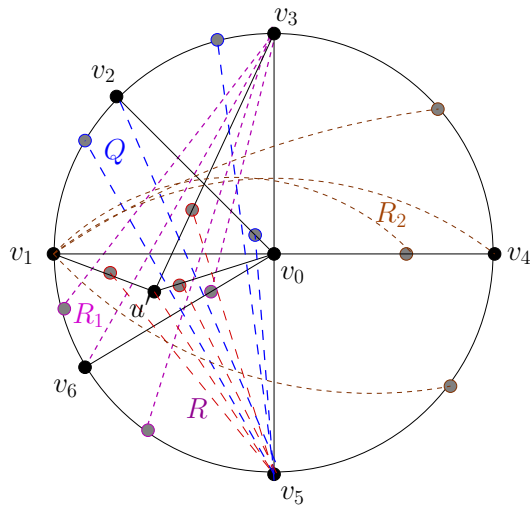}
\caption[Starting graph for search example]{Starting graph for search example: case (b)(i) 1.1.1.1.1 in Theorem 18 of \cite{Robinson09}.}
\label{searcheg_start}
\end{center}
\end{figure}

We now give an example of how \texttt{exception\_generator} is used in proofs. In the main result (Theorem 18) of \cite{Robinson09}, case (b)(i) 1.1.1.1.1, we start with the graph of Figure \ref{searcheg_start}.

Note that the edges marked $Q$, $R$, $R_{1}$, and $R_{2}$ in this graph each have four possible placements in the graph (represented by dotted lines in Figure \ref{searcheg_start}). Thus, there are in fact $4^{4}$ possible starting graphs $G_{i}$. For each of these graphs $G_{i}$, we consider the set $S_{3} = \{v_{0}, v_{1}, v_{3}, v_{5}\}$, with the aim of discovering whether some path $P$ can be added to $G_{i}$ such that $S_{3}$ is not a separating set of $G_{i}\cup P$, and $G_{i} \cup P$ does not contain a $W_{7}$-subdivision. The \texttt{exception\_generator} function can be used as follows:

\begin{itemize}
\item[] For each starting graph $G_{i}$:
	\begin{itemize}
	\item[] Let $A$, $B$ be the two components of $G_{i} - S_{3}$
	\item[] Call \texttt{exception\_generator($G_{i}$, $A$, $|V(A)|$, $B$, $|V(B)|$)}
	\end{itemize}
\end{itemize}

Running this algorithm finds that each generated graph contains a $W_{7}$-subdivision. Thus, it can be assumed from this point onwards in the proof that $S_{3}$ is a separating set.

\section{Algorithm for finding $W_{k}$-subdivisions}
\label{wk_alg}

The main algorithm in this section is \texttt{findkwheel}, and is given in Section \ref{findkwheel}. It solves SHP($W_{k}$) for any given input graph, and for any value of $k \ge 4$. This algorithm runs in exponential time, but performs sufficiently quickly on input graphs of small size.

The algorithm \texttt{findkwheel} makes a call to another algorithm, \texttt{iskwheel}, which determines whether or not some input graph $G$ is a $W_{k}$-subdivision, for a given value of $k \ge 4$.

\subsection{\texttt{iskwheel}}
\label{iskwheel}

\texttt{iskwheel(G, k)} takes two arguments, a graph $G$ and an integer $k$, and determines whether or not $G$ is a $W_{k}$-subdivision.

A 2-connected graph $G$ is isomorphic to the wheel $W_{k}$ if the following is true:
\begin{itemize}
\item $|V(G)| = k + 1$;
\item $G$ contains exactly $k$ vertices of degree 3; and
\item $G$ contains exactly one vertex of degree $k$.
\end{itemize}

A graph $G$ is a $W_{k}$-subdivision if, after contracting all vertices of degree 2, $G$ becomes isomorphic to the graph $W_{k}$.

Thus, the function \texttt{iskwheel} uses the following algorithm:

\begin{itemize}
\item[] Step 1. Check to see if $G$ is two-connected. If not, $G$ cannot be a $W_{k}$-subdivision: return null.
\item[] Step 2. Contract all vertices of degree 2 in $G$.
\item[] Step 3. If $G$ contains exactly $k + 1$ vertices, $k$ of which have degree 3, and one of which has degree $k$, then return $G$; otherwise return null.
\end{itemize}

Determining if $G$ is two-connected in Step 1 is done with a worst-case complexity of $O(n^{2})$, using an implementation of Hopcroft's biconnectivity algorithm \cite{AhoHopUll}. (The implementation is given in Appendix \ref{app_is2connected}.) Contracting all vertices of degree 2 until there are no such vertices left has a complexity of $O(n)$. Counting the degrees of remaining vertices in Step 3 is $O(n)$. Thus, the entire algorithm's complexity is $O(n^{2})$.

The exact code is given in Appendix \ref{app_iskwheel}.

\subsection{\texttt{findkwheel}}
\label{findkwheel}

The function \texttt{findkwheel(G, k)} also takes as its arguments a graph $G$ and an integer $k$. This function searches for a $W_{k}$-subdivision as a subgraph of $G$; if such a subgraph exists, \texttt{findkwheel} will return it, otherwise it returns null. This is done by recursively testing all subgraphs obtained by removing a single edge from the input graph. Base cases are graphs that are $W_{k}$-subdivisions, or small graphs that clearly do not contain such a subdivision.

The following algorithm is used.

\begin{itemize}
\item[] Step 1. Remove any vertices in $G$ with degree zero. 
\item[] Step 2. Call \verb|iskwheel(G)|. If $G$ is not a $W_{k}$-subdivision, go to Step 3; otherwise return $G$. 
\item[] Step 3. If $|V(G)| < k+1$, or $|E(G)| < 2k$, then $G$ is too small to contain a $W_{k}$-subdivision. Return null.
\item[] Step 4. If $G$ contains no vertex with degree $\ge k$, return null. 
\item[] Step 5. For each edge $e$ that exists, call \texttt{findkwheel(G - e, k)}. If a $W_{k}$-subdivision is found, return that graph, otherwise continue to Step 6. 
\item[] Step 6. $G$ does not contain a $W_{k}$-subdivision. Return null.
\end{itemize}


This algorithm runs in exponential time, but still performs effectively on reasonably small graphs. The code is given in Appendix \ref{app_findkwheel}.

\section{Concluding remarks}

The proofs of the main results in \cite{Robinson08} and \cite{Robinson09} regarding $W_{6}$- and $W_{7}$-subdivisions are of sufficient complexity that completing such proofs without the aid of a computer program becomes extremely difficult. The algorithms presented in this paper, particularly the \texttt{exception\_generator} algorithm given in Section \ref{furthertests}, form a key component in automating the generation and testing of graphs required as test cases in these proofs. The \texttt{exception\_generator} algorithm may well be useful in developing other characterizations of SHP-related problems, where a similar approach is adopted in the proof of moving from a problem with a good characterization to one without.

\begin{appendices}

\section{C code for wheelproof}
\label{app_wheelproof}

\begin{verbatim}
/* Takes an integer k, and outputs all exception graphs - graphs that
** do not contain a W_{k}-subdivision - from the starting point of the
** proof.
*/
static Graphlist *wheelproof(const int k)
{
   Graph *graph;
   Graphlist *exception_list = (Graphlist *) malloc(sizeof(Graphlist));
   Graph *prev = NULL;
   Graph *next_exception;
   int i, j, l, m, x, y;
   int u = k;
   int u1 = k+1;
   int u2 = k+2;
	
   exception_list->size = 0;
	
   /* Make the starting graph W_{k-1}. */
   graph = makewk(graph, k-1);
	
   /* Adding vertex u and joining to the centre vertex */
   graph = addvertex(graph, u);
   graph = addedge(graph, 0, u);


/* Main loop: creates edges between u and u1, and between u and u2,
** for each possible placement of u1 and u2. Each resulting graph
** is tested for a W_{k}-subdivision.
*/

   /* For each vertex i in the graph (except where i=u): */
   for (i = 0; i <= graph->highestid; i++)
   {
      if (i != u && graph->vertices[i] != NULL)
      {
         if (i != 0) /* Vertex 0 is already adjacent to u */
         {
            /* where u1 is an already existing vertex - add edge. */
            graph = addedge(graph, u, i);

            /* Now look at possibilities for u2. */
            for (l = i; l <= graph->highestid; l++)
            {
               if (l != u && graph->vertices[l] != NULL)
               {
                  /* where u2 is already existing vertex: */
                  if (l != 0)
                  {
                     graph = addedge(graph, u, l);
                     
                     /* Check if the two new edges make a W_{k} subdivision */
                     if (!findkwheel(graph, k, 0, 0))
                     {
                        if (!is3connected(graph))
                           printf("Graph not 3-connected.\n");
                        else
                        {
                           printf("Exception found.\n");
                           next_exception = graphcpy(graph);
                           if (prev != NULL)
                              prev->next = next_exception;
                           else
                              exception_list->head = next_exception;
                           prev = next_exception;
                           exception_list->size++;
                        }
                     }
                     
                     /* Remove u - u2 edge again. */
                     graph = removeedge(graph, u, l);
                  }

                  /* This time, u2 is a new vertex on some 'edge'
                  ** (since each edge represents a path in G).
                  */
                  for (m = i; m < graph->vertices[l]->degree; m++)
                  {
                     y = graph->vertices[l]->neighbours[m];
                     if (y > l && y != u && graph->vertices[y] != NULL)
                     {
                        graph = addvertex(graph, u2);
                        graph = addedge(graph, u, u2);
                        graph = expand_edge(graph, l, y, u2);
                        
                        /* Check if the two new edges make a W_{k} subdivision */
                        if (!findkwheel(graph, k, 0, 0))
                        {
                           if (!is3connected(graph))
                              printf("Graph not 3-connected.\n");
                           else
                           {
                              printf("Exception found:\n");
                              next_exception = graphcpy(graph);
                              if (prev != NULL)
                                 prev->next = next_exception;
                              else
                                 exception_list->head = next_exception;
                              prev = next_exception;
                              exception_list->size++;
                           }
                        }
                        
                        /* Remove u2 again. */
                        graph = removeedge(graph, u, u2);
                        graph = contractvertex(graph, u2);
                     }
                  }
               }
            }
         
            /* Remove the edge that was added before trying next possibility. */
            graph = removeedge(graph, u, i);
         }

         /* Look at all possibilities for u1 where u1 is a new vertex
         ** that lies on the path between i and one of its 'neighbours'
         */         
         for (j = 0; j < graph->vertices[i]->degree; j++)
         {
            x = graph->vertices[i]->neighbours[j];
            if (x > i && x != u && graph->vertices[x] != NULL)
            {
               /* New edge joins at a new vertex which splits an
               ** existing edge into two edges. */
               graph = addvertex(graph, u1);
               graph = addedge(graph, u, u1);
               graph = expand_edge(graph, i, x, u1);
               
               /* Now select u2 */
               for (l = i; l <= graph->highestid; l++)
               {
                  if (l != u && graph->vertices[l] != NULL)
                  {

                     /* u2 is an already existing vertex */
                     if (l != 0)
                     {
                        graph = addedge(graph, u, l);
                        /* Check if the two new edges make a W_{k} subdivision */
                        if (!findkwheel(graph, k, 0, 0))
                        {
                           if (!is3connected(graph))
                              printf("Graph not 3-connected.\n");
                           else
                           {
                              printf("Exception found:\n");
                              next_exception = graphcpy(graph);
                              if (prev != NULL)
                                 prev->next = next_exception;
                              else
                                 exception_list->head = next_exception;
                              prev = next_exception;
                              exception_list->size++;
                           }
                        }
                        
                        /* Remove u - u2 edge again. */
                        graph = removeedge(graph, u, l);
                     }
                     
                     /* u2 is a new vertex: generate all possibilities */
                     for (m = i; m < graph->vertices[l]->degree; m++)
                     {
                        y = graph->vertices[l]->neighbours[m];
                        if (y > l && y != u && graph->vertices[y] != NULL)
                        {
                           graph = addvertex(graph, u2);
                           graph = addedge(graph, u, u2);
                           graph = expand_edge(graph, l, y, u2);
                           
                           /* Check if the two new edges make a W_{k} subdivision */
                           if (!findkwheel(graph, k, 0, 0))
                           {
                              if (!is3connected(graph))
                                 printf("Graph not 3-connected.\n");
                              else
                              {
                                 printf("Exception found:\n");
                                 next_exception = graphcpy(graph);
                                 if (prev != NULL)
                                    prev->next = next_exception;
                                 else
                                    exception_list->head = next_exception;
                                 prev = next_exception;
                                 exception_list->size++;
                              }
                           }
                           
                           /* Remove u2 again. */
                           graph = removeedge(graph, u, u2);
                           graph = contractvertex(graph, u2);
                        }
                     }
                  }
               }
               
               /* Take u1 out again. */
               graph = removeedge(graph, u, u1);
               graph = contractvertex(graph, u1);
            }
         }
      }
   }
   
   graph = removeedge(graph, 0, u);
   graph = removevertex(graph, u);
   
   return exception_list;
}

/* makewk: returns the graph W_{k} for given k */
static Graph *makewk(Graph *graph, int k) 
{
      int i = 0;
       graph = initialise_graph(graph);
      
      /* Create k+1 vertices */
      while (i <= k)
      {
         graph = addvertex(graph, i);
         i++;
      }
      
      i = 1;
      
      /* Create k spokes */
      while (i <= k)
      {
         graph = addedge(graph, 0, i);
         i++;
      }
      
      i = 1;
      
      /* Create rim of wheel */
      while (i < k)
      {
         graph = addedge(graph, i, i+1);
         i++;     
      }
      
      graph = addedge(graph, k, 1);
      return graph;
}
\end{verbatim}

\section{C code for exception\_generator}
\label{app_exception_generator}

\begin{verbatim}
/* Function to process all the possible exceptions that can be
** generated from each starting graph.
*/
static void exception_generator(Graph *graph, int sectionA[], int asize,
   int sectionB[], int bsize)
{
   int i, j, k, l, p, p1, n, m, skip=0, skip1=0;
   Vertex *currvertex, *currvertex1;
   int newvertex1 = graph->highestid;
   int newvertex2 = (graph->highestid) + 1;
   int nbr[MAXDEGREE], nbr1[MAXDEGREE];

/* Process possible graphs */

   for (i=0; i<asize; i++)
   {
      for (j=0; j<bsize; j++)
      {
         /* add new path: endpoints are vertices that already exist */
         graph = addedge(graph, sectionA[i], sectionB[j]);

         /* Is there a W7? */
         if (findkwheel(graph, 7, 0, 0) == NULL)
         {
            printf("Exception:\n");
            printgraph(graph);
         }
         /* remove new path */
         graph = removeedge(graph, sectionA[i], sectionB[j]);

         /* add new path: endpoints are new vertex in section A
         ** and already existing vertex in section B
         */         
         n = 0;
         currvertex = graph->vertices[sectionA[i]];
         while (n < currvertex->degree)
         {
            nbr[n] = currvertex->neighbours[n];
            n++;
         }

         /* For each neighbour k of i, try expanding the edge ik */
         for (k=0; k<n; k++)
         {

            /* If we've looked at this neighbour before, skip it. */
            for (p=0; p<i; p++)
            {
               if (sectionA[p] == nbr[k])
                  skip = 1;
               else
                  skip = 0;
            }
			
            /* ... otherwise, create a new vertex in section A along the
            ** path between i and k, and make a path between this and
            ** vertex j in section B
            */
            if (!skip)
            {
               graph = addvertex(graph, newvertex1);
               graph = expand_edge(graph, sectionA[i], nbr[k], newvertex1);
               graph = addedge(graph, newvertex1, sectionB[j]);
               if (findkwheel(graph, 7, 0, 0) == NULL)
               {
                  printf("Exception:\n");
                  printgraph(graph);
               }

               /* Remove path. */
               graph = removeedge(graph, newvertex1, sectionB[j]);
               graph = contractvertex(graph, newvertex1);
            }
         }

         /* add new path: endpoints are new vertex in section B
         ** and already existing vertex in section A
         */         
         n = 0;
         currvertex = graph->vertices[sectionB[j]];
         while (n < currvertex->degree)
         {
            nbr[n] = currvertex->neighbours[n];
            n++;
         }

         /* For each neighbour k of j, try expanding the edge jk */
         for (k=0; k<n; k++)
         {

            /* If we've looked at this neighbour before, skip it. */
            for (p=0; p<j; p++)
            {
               if (sectionB[p] == nbr[k])
                  skip = 1;
               else
                  skip = 0;
            }

            /* ... otherwise, create a new vertex in section B along the
            ** path between j and k, and make a path between this and
            ** vertex i in section A
            */
            if (!skip)
            {
               graph = addvertex(graph, newvertex1);
               graph = expand_edge(graph, sectionB[j], nbr[k], newvertex1);
               graph = addedge(graph, newvertex1, sectionA[i]);
               if (findkwheel(graph, 7, 0, 0) == NULL)
               {
                  printf("Exception:\n");
                  printgraph(graph);
               }
               graph = removeedge(graph, newvertex1, sectionA[i]);
   
               /* don't contract new vertex yet, but rather... */
   
            /* add new path: endpoints are new vertex in section B
            ** (that is, the one we just made) and new vertex in section A
            */         
            m = 0;
            currvertex1 = graph->vertices[sectionA[i]];
            while (m < currvertex1->degree)
            {
               nbr1[m] = currvertex1->neighbours[m];
               m++;
            }

            /* For each neighbour l of i, try expanding the edge il */
            for (l=0; l<m; l++)
            {

                  /* If we've looked at this neighbour before, skip it. */
                  for (p1=0; p1<i; p1++)
                  {
                     if (sectionA[p1] == nbr1[l])
                        skip1 = 1;
                     else
                        skip1 = 0;
                  }

               /* ... otherwise, create a new vertex in section A along the
               ** path between i and l, and make a path between this and
               ** the new vertex (newvertex1) in section B
               */
                  if (!skip1)
                  {
                     graph = addvertex(graph, newvertex2);
                     graph = expand_edge(graph, sectionA[i], nbr1[l], newvertex2);
                     graph = addedge(graph, newvertex1, newvertex2);
                     if (findkwheel(graph, 7, 0, 0) == NULL)
                     {
                        printf("Exception:\n");
                        printgraph(graph);
                     }
                     graph = removeedge(graph, newvertex1, newvertex2);
                     graph = contractvertex(graph, newvertex2);
                  }
               }
               graph = contractvertex(graph, newvertex1);
            }
         }
      }
   }
}
\end{verbatim}

\section{C code for testing biconnectivity of a graph}
\label{app_is2connected}

\begin{verbatim}
/* is2connected: Returns 1 if the graph starting at input vertex
** 'head' is 2-connected. Returns 0 otherwise.
*/
int is2connected(Graph *graph)
{
   int visited[MAXGRAPHSIZE];
   int dfnumber[MAXGRAPHSIZE];
   int low[MAXGRAPHSIZE];
   int father[MAXGRAPHSIZE];
   int count = 0;
   int i = 0;

   for (i=0;i<MAXGRAPHSIZE;i++)
   {
      visited[i] = 0;
      dfnumber[i] = -1;
      low[i] = -1;
      father[i] = -1;
   }

/* Find the first vertex in the graph. */
   i = 0;
   while (graph->vertices[i] == NULL)
      i++;

   return (is2conn_rec(graph, i, visited, dfnumber, low, father, &count));
}

/* is2conn_rec: Recursive function used by is2connected. */
static int is2conn_rec(Graph *graph, int v_id, int visited[], int dfnumber[],
   int low[], int father[], int *count)
{
   Vertex *v = graph->vertices[v_id];
   Vertex *w;
   int w_id;
   int i = 0;

   visited[v_id] = TRUE;
   dfnumber[v_id] = *count;
   (*count)++;
   low[v_id] = dfnumber[v_id];

   while (i < v->degree)
   {
      w_id = v->neighbours[i];
      w = graph->vertices[w_id];
      if (w == NULL)
      {
         printf("Error: vertex connected to vertex that doesn't exist.\n");
         exit(1);
      }
      if (visited[w_id] == FALSE)
      {
         father[w_id] = v_id;
         if (!is2conn_rec(graph, w_id, visited, dfnumber, low, father, count))
            return FALSE;
         if (low[w_id] >= dfnumber[v_id] && ((dfnumber[v_id] != 0) || i > 0) )
            return FALSE; 
         low[v_id] = min(low[v_id], low[w_id]);
      }
      else if (father[v_id] != w_id)
      {
         low[v_id] = min(low[v_id], dfnumber[w_id]);
      }
      i++;
   }

   return TRUE;
}
\end{verbatim}

\section{C code for iskwheel}
\label{app_iskwheel}

\begin{verbatim}
/* iskwheel: Takes a graph 'graph' and an integer k. If the input 
** graph is the graph W_{k} once all vertices of degree 2 have been
** contracted, then the function returns a copy of the input graph 
** with all such vertices contracted.
** If the input graph is not the graph W_{k}, the function returns
** the null pointer.
*/

Graph *iskwheel(Graph *graph, int k)
{
   int countk = 0;
   int count3 = 0;
   int countcontracted = 1;
   int i=0;
   Vertex *v; 
   Graph *newgraph;

   if (!is2connected(graph)) return NULL; /* Graph must be 2-connected. */
   newgraph = graphcpy(graph);

/* For each vertex in the graph: if v is of degree 2, contract v */

   while (countcontracted != 0)
   {
      i = 0;
      countcontracted = 0;
      while (i <= newgraph->highestid)
      {
         v = newgraph->vertices[i];
         if (v != NULL)
         {
            if (v->degree == 2)
            {
               newgraph = contractvertex(newgraph, i);
               countcontracted++;
            }
         }
         i++;
      }
   }

/* If v is degree 3, increment counter of degree 3 vertices
** If v is of degree k, increment counter of degree k vertices
*/
   i = 0;
   while (i <= newgraph->highestid)
   {
      v = newgraph->vertices[i];
      if (v != NULL)
      {
         if (v->degree == 3)
            count3++;
         else if (v->degree == k)
            countk++;
      }
         i++;
   }
	
/* Number of vertices (not including those of degree 2 which
** were contracted) must equal k+1 for graph to be W_{k}-subdivision.
*/
   if (newgraph->size != k+1)
   {
      killgraph(newgraph);
      return NULL;
   }

/* Must be k vertices of degree 3 and 1 vertex of degree k. */

   if (count3 == k && countk == 1)
      return newgraph;

/* Special case for W_{3}, where there are 4 vertices of degree 3. */

   else if (k == 3 && count3 == 4)
      return newgraph;

/* If graph is not a W_{k}-subdivision, return NULL. */
   else
   {
      killgraph(newgraph);
      return NULL;
   }
}
\end{verbatim}

\section{C code for findkwheel}
\label{app_findkwheel}

\begin{verbatim}
/* findkwheel: Takes as input a graph and an integer k. If
** the input graph contains a W_{k} subdivision, the function
** returns the input graph contracted to be W_{k}. Otherwise,
** the null pointer is returned.
*/

Graph *findkwheel(Graph *graph, int k, int startvertex1, int startvertex2)
{
   Graph *subgraph, *foundwheel;
   Vertex *v;
   int i, j;

/* Any vertices of degree 0 are removed from the graph.
** This is so they do not affect the vertex count in
** iskwheel().
*/

   i = 0;
   while (i <= graph->highestid)
   {
      v = graph->vertices[i];
      if (v != NULL)
      {
         if (v->degree == 0)
         {
            graph = removevertex(graph, i);
         }
      }
      i++;
   }

/* If the graph is not W_{k}, the function is called recursively
** to test the removal of every possible combination of edges
** between vertices higher than startvertex1 and startvertex2.
** If some combination of edge removal results in W_{k}, the program
** exits the while loop and returns successfully.
** Once there are too few edges or vertices left in the graph, or 
** once all edge removal possibilities have been tried, the function
** returns NULL.
*/
   if ((subgraph = iskwheel(graph, k)) == NULL)
      /* only do this if the graph isn't W_{k} */
   {
	
   /* Too few edges to be able to remove any more,
   ** or too few vertices to make W_{k} */
      if (graph->edges <= 2*k || graph->size < k+1)
         return NULL;
   /* or no vertex left of degree at least k */
      for (i = 0; i <= graph->highestid; i++)
      {
         if (graph->vertices[i] != NULL && graph->vertices[i]->degree >= k)
            break;
      }
      if (i > graph->highestid)
         return NULL;

      i = startvertex1;
   /* Removing each edge in turn: */
      while (i <= graph->highestid) 
      {
         if (i == startvertex1) // first time through outer loop
            j = max(i, startvertex2);
         else
            j = i;
         while (j <= graph->highestid)
         {
            if (edgeexists(graph,i,j) && j > i)
               /* j > i check ensures only edges in one direction are detected */
            {
               subgraph = graphcpy(graph);
   /* Call the function recursively on the graph with one fewer edge. */
               if ((foundwheel = findkwheel(removeedge(subgraph,i,j), k, i, j)) != NULL)
               {
                  return foundwheel;
               }
               killgraph(subgraph);
            }
            j++;
         }
         i++;
      }
      return NULL;
   }
   return subgraph;
}
\end{verbatim}

\end{appendices}


\begin{thebibliography}{1}

\bibitem{AhoHopUll}
Alfred~V. Aho, John~E. Hopcroft, and Jeffrey~D. Ullman.
\newblock {\em The Design and Analysis of Computer Algorithms}.
\newblock Addison-Wesley, Reading, MA, 1974.

\bibitem{Farr88}
G.~Farr.
\newblock The subgraph homeomorphism problem for small wheels.
\newblock {\em Discrete Math.}, 71:129--142, 1988.

\bibitem{G&J79}
M.R. Garey and D.S. Johnson.
\newblock {\em Computers and Intractability: A Guide to the Theory of
  NP-Completeness}.
\newblock W. H. Freeman, New York, 1979.

\bibitem{R&S95}
N.~Robertson and P.D. Seymour.
\newblock Graph minors. {XIII}. {T}he disjoint paths problem.
\newblock {\em J. Combin. Theory Ser. B}, 63:65--110, 1995.

\bibitem{Robinson08}
Rebecca Robinson and Graham Farr.
\newblock Structure and recognition of graphs with no 6-wheel subdivision.
\newblock Published online in \emph{Algorithmica}, January 2008; awaiting print
  publication.

\bibitem{Robinson09}
Rebecca Robinson and Graham Farr.
\newblock Graphs with no 7-wheel subdivision.
\newblock Technical Report 2009/239, Clayton School of Information Technology,
  Monash University (Clayton Campus), 2009.

\end{thebibliography}
\end{document}